\documentclass[11pt]{article}
\usepackage{amsmath,amssymb}
\usepackage[dvips]{graphicx}
\usepackage[abs]{overpic}
\newcommand{\ba}{\begin{alignat}{3}}

\begin{document}


\begin{flushright}
OU-HET 624
\\
February 20, 2009
\end{flushright}
\vskip3cm
\begin{center}
{\LARGE {\bf Holographic RG flow dual to attractor flow \\[0.2cm] in extremal black holes}}
\vskip3cm
{\large 
{\bf Kyosuke Hotta\footnote{hotta@het.phys.sci.osaka-u.ac.jp} 
}
}

\vskip1cm
{\it Department of Physics, Graduate School of Science, 
\\
Osaka University, Toyonaka, Osaka 560-0043, Japan}
\end{center}

\vskip1cm
\begin{abstract}
We extend the discussion of the ``Kerr/CFT correspondence" and its recent developments to the more general gauge/gravity correspondence in the full extremal black hole space-time of the bulk by using a technique of the holographic renormalization group (RG) flow.
It is conjectured that the extremal black hole space-time is holographically dual to the chiral two dimensional field theory. 
Our example is a typical four dimensional Reissner-Nordstrom black hole, a system in which the M5-brane is wrapped on  four cycles of Calabi-Yau threefold.
In five dimensional supergravity view point this near horizon geometry is $\textrm{AdS}_3\times \textrm{S}^2$, and three dimensional gravity coupled to moduli fields is effectively obtained after a dimensional reduction on $\textrm{S}^2$. 
Constructing the Hamilton-Jacobi equation, we define the holographic RG flow from the three dimensional gravity.
The central charge of the Virasoro algebra is calculable from the conformal anomaly at the point where the beta function defined from gravity side becomes zero.
In general, we can also identify the c-function of the dual two dimensional field theory.
We show that these flow equations are completely equivalent to  not only BPS but also non-BPS attractor flow equations of the muduli fields.
The attractor mechanism by which the values of the moduli fields are fixed at the event horizon of the extremal black hole can be understood equivalently to  the fact that the RG flows are fixed at the critical points in the dual field theory.
\end{abstract}


\vfill\eject


\section{Introduction}
\label{intro}

The $\textrm{AdS}/\textrm{CFT}$ or a more general claim, the gauge/gravity  correspondence~\cite{m,gkp,witten0} gives some insights  to the whole picture of quantum theory of gravity, in particular, the understanding of black hole entropy from microscopic view point. 
A recent work, the ``$\textrm{Kerr}/\textrm{CFT}$ correspondence"~\cite{ghss}, is one further evidence of them.
Combining it with the later developments~\cite{hhknt}-\cite{cmn}, one can state as follows: as far as extremal black holes are considered, there exists chiral two dimensional conformal field theory ($\textrm{CFT}_2$) satisfying the Virasoro algebra on a boundary of (warped) $\textrm{AdS}_3$ at the event horizon.\footnote{See also refs.~\cite{car}-\cite{kkp} on  further discussions of the Virasoro algebra at the horizon.} 
The purpose of the present paper is to extend these discussions to the full  black hole space-time in the bulk and confirm the gauge/gravity correspondence between the extremal black hole and the quantum field theory ($\textrm{QFT}$) on the boundary.

An important example constructed in~\cite{hhknt} is an extension of the Ba\~{n}ados-Teitelboim-Zaneli (BTZ) black hole~\cite{btz} with a non-trivial potential, and has the $\textrm{AdS}_3$ geometries only at the spatial infinity and the event horizon.
With a similar prescription in 3d gravity to that done by Brown and Henneaux~\cite{brown, ghss}, one can explicitly verify  that there exists the $\textrm{CFT}_2$ satisfying the Virasoro algebras on both boundaries,\footnote{See also refs.~\cite{cmp} for   a 3d black hole which interpolates two $\textrm{AdS}_2$.}
and furthermore  they are surely connected by the ``holographic" RG flow~\cite{fgpw,st}.
If we relate the radial coordinate and scalars in the bulk gravity to the scale and running couplings, respectively, of the corresponding $\textrm{QFT}$ on the boundary, the Hamilton-Jacobi equation of gravity theory can be seen as the RG flow equation of boundary theory~\cite{dbvv}.\footnote{
See also previous works  based on this formalism~\cite{noo1}-\cite{fms1}. }

We can consider that there is also such a  gauge/gravity correspondence between the effective 3d gravity and the boundary $\textrm{QFT}_2$ for more general, higher dimensional extremal black objects, such as the  4d Reissner-Nordstrom (RN) black hole, because there seems to exist the $\textrm{AdS}_3/\textrm{CFT}_2$ correspondence, at least at the horizon.
In the  4d  extremal RN black hole, if one assumes that the  4d gauge field comes from Kaluza-Klein (KK) $U(1)$ part of the compact $\text{S}^1$, one can realize the (warped) $\textrm{AdS}_3$ geometry at the horizon, and then one of the Virasoro algebras is derived~\cite{aot1,hmns}. 

According to this observation, as the easiest example, we consider a system  in which the M5-brane is wrapped on  four cycles of Calabi-Yau threefold ($\textrm{CY}_3$).
This gives the 4d RN black hole with moduli fields, but the electric field can be regarded as  the KK momentum of the compact $\text{S}^1$.
So it is equivalent to a 5d rotating black ring solution with magnetic fluxes, or the $\text{D0}$-$\text{D4}$ system in type IIA view point.
Its near horizon geometry is $\textrm{AdS}_3\times \textrm{S}^2$, and it is easy to obtain decoupled 3d gravity with scalars when one compactifies the 5d gravity on $\text{S}^2$.
We investigate the gauge/gravity correspondence between this effective 3d gravity and the $\textrm{QFT}_2$ on the boundary.

It is expected to  derive the Virasoro algebra dual to $\textrm{AdS}_3$ just at the horizon by a similar calculation to that done in refs.~\cite{ghss}, but this paper devotes to see the whole behavior of the flow of the dual $\textrm{QFT}_2$ through the Hamilton-Jacobi formalism.
By a  direct application of the formalism to our effective 3d gravity of the M5-system, we calculate the central charge of the Virasoro algebra from the conformal anomaly at the critical point where the beta function defined from gravity side becomes zero, correspondingly to the $\textrm{AdS}_3$ geometry at the horizon.
The microscopic black hole entropy can be  realized  without directly deriving the Virasoro algebra.
Furthermore, it is also possible  to identify Zamolodchikov's c-function~\cite{z} from gravity side.
These  results conclude the consistent extention of the horizon $\textrm{CFT}_2$ to the non-conformal $\textrm{QFT}_2$ in the full black hole space-time.
For more general charge configurations, it is of course difficult to obtain the decoupled 3d gravity and to see such a direct  correspondence between 3d gravity in the bulk and $\textrm{QFT}_2$ on the boundary.
However, we here conjecture that the extremal black hole space-time  is holographically dual to the chiral $\textrm{QFT}_2$.

This claim is motivated by  attractor behaviors of scalar fields in  extremal black holes~\cite{fks,fk,attst}.
In extremal black holes with gauge fields and scalars,  values of the scalars at the horizon are completely determined by the black hole charges, irrespective of  the values at the spatial infinity.
Moreover, the geometry of the extremal black hole generally becomes $\textrm{AdS}$ at the horizon~\cite{klr, astefa}.
This attractor mechanism was first observed in supersymmetric black hole solutions~\cite{fks}-\cite{attst},~\cite{gkk}-\cite{denef}, and later extended to non-supersymmetric ones~\cite{gijt,tt}.

It has been said loosely that this attractor phenomenon resembles a condition that RG flows in $\textrm{QFT}$ are fixed at the critical   point.\footnote{There are a lot of recent discussions on the attractor mechanism~\cite{sen}-\cite{psrv}.
Especially, a relation between the attractor mechanism and the c-function has been pointed out in refs.~\cite{gjmt,anyy}, and  it was also discussed that there are first order attractor flow equations even for non-BPS black holes in refs.~\cite{cersole,aaot,psrv}.}
Then in this paper we show, in a more convincing way, that  not only BPS but also  non-BPS attractor flow equations of the M5-system are actually equivalent to the holographic RG flow equations obtained by the Hamilton-Jacobi formalism.
The effective 3d view point is very useful to see their connection,
independently of supersymmetry of bulk theory.
And it is natural to expect to have either BPS or non-BPS first order flow equations when one uses the canonical formalism.
Our result presents an explicit evidence of the equivalence between the  attractor flow  and the holographic RG flow.
And it confirms that the attractor mechanism can holographically understood as the fact that the RG flow of the dual $\textrm{QFT}_2$ are always attracted at the IR fixed point since the scalars have to be identified as the running couplings in the formalism.
Therefore, it can be claimed that the attractor behaviors appearing  in extremal black holes, in general,  indicates the existence of the dual $\textrm{QFT}$ along the holographic RG flows.

Our paper is organized as follow: 
in section \ref{5d4d} we see a connection between 5d and 4d supergravity theory with eight supercharges and show the BPS attractor flow equations.
In section \ref{3d} the M5-system is considered as the easiest example, and the dimensional reduction on $\text{S}^2$ is carried out in order to obtain the effective 3d  theory.
Then, after reviewing the Hamilton-Jacobi formalism in subsection \ref{reviewofhj}, we identify the beta function and the c-function of the M5-system from the effective 3d gravity theory in subsection \ref{betac}.
In section \ref{attractor} we relate these flows to the BPS attractor flow equations and also derive non-BPS ones as an application of our result.
The concluding remarks are described  in section \ref{conclusion}.
In appendix \ref{superconformal} we note 5d $N=2$ superconformal  gravity so as  to deal with on-shell 5d supergravity action.
We show some detailed calculations for solving 4d equations of motion  in  appendix \ref{bhpotential}, and for the Hamilton-Jacobi formalism in appendix \ref{correct}.


\section{Very Special Geometry and Special Geometry}
\label{5d4d}

In this section we review a relation between 5d  and 4d supergravity with eight supercharges~\cite{gst}.
In this setup, typical 4d RN black holes are embedded into 5d gravity.
In section \ref{rg} the boundary $\textrm{QFT}_2$ will be found as the  holographic dual to the effective 3d gravity obtained  from  this 5d gravity.
See useful reviews~\cite{5dreview} and~\cite{mohauptreview} for 5d and 4d $N=2$ supergravity, respectively.

Let us first consider the 5d $N=2$ supergravity action which comes from  11d supergravity compactified on a $\textrm{CY}_3$.
We write $M^a$ and $c_{abc}$ as the K\"{a}hler moduli and the intersection number of the $\textrm{CY}_3$, respectively, where $a,b,\cdots=(1,2,\cdots h^{1,1})$.
Then the action is expressed as 
\begin{align}
\mathcal{I}_{(5)}=\frac{1}{4\pi^2}\int 
d^5x
\sqrt{-g_{(5)}}
\Bigg[
R^{(5)}
&-\frac{1}{2}\left(\frac{2\mathcal{N}_a\mathcal{N}_b}{3\mathcal{N}^2}
-\frac{\mathcal{N}_{ab}}{\mathcal{N}}\right)\partial_M M^a\partial^M M^b\notag\\
&\,\,\,\,\,\,\,\,\,\,\,\,\,\,\,\,\,\,\,\,\,\,\,\,\,
-\frac{\mathcal{N}^{2/3}}{2}G_{ab}F^a_{MN}F^{bMN}\Bigg]+\mathcal{I}_{CS},
\label{5daction2}
\end{align}
\begin{equation}
\mathcal{I}_{CS}=\frac{1}{4\pi^2}\int\frac{1}{6}\,c_{abc}\,A^a\wedge F^b\wedge F^c.
\end{equation}
The indices $M, N, \cdots$ label the 5d space-time coordinate.
The functions $\mathcal{N}$, $\mathcal{N}_a$, $\mathcal{N}_{ab}$ and $G_{ab}$ are defined by
\begin{equation}
\mathcal{N}=\frac{1}{6}c_{abc}M^{a}M^{b}M^{c},\,\,\,\,\,\,\,\,
\mathcal{N}_{a}=\frac{1}{2}c_{abc}M^{b}M^{c},\,\,\,\,\,\,\,\,
\mathcal{N}_{ab}=c_{abc}M^{c},
\end{equation}
\begin{equation}
G_{ab}=\frac{1}{2}\left(\frac{\mathcal{N}_a\mathcal{N}_b}{\mathcal{N}^2}
-\frac{\mathcal{N}_{ab}}{\mathcal{N}}\right).
\end{equation}
Because we can fix the value of $\mathcal{N}$ which can be seen as a total volume of the $\textrm{CY}_3$, a constraint  
$\mathcal{N}=1$
is often imposed due to the decoupling of hypermultiplets.
This theory is known as {\it very special geometry}. 
As is noted in appendix \ref{superconformal}, when we start with 5d superconformal gravity, $\mathcal{N}=1$ is automatically derived from an equation of motion of an auxiliary field.
Alternatively, we can assume that it is derived as a solution  from the action (\ref{5daction2}) as well.

Here let us carry out a compactification on $\text{S}^1$.
Decomposing the 5d metric and $U(1)$ gauge field like
\begin{align}
ds^2_{(5)}&=e^{-s}(g^{(4)}_{mn}dx^mdx^n)+e^{2s}(dy+A_mdx^m)^2,\notag\\
A^a&=A^a_mdx^m-a^a(dy+A_mdx^m),
\label{5dto4d}
\end{align}
we have the 4d supergravity action
\begin{align}
\mathcal{I}_{(4)}=\frac{1}{2\pi}\int d^4x\sqrt{-g_{(4)}}
\Big(
R^{(4)}-2G_{a\overline{b}}\partial_m z^a\partial^m \overline{z}^{\overline{b}}
-\frac{1}{4}\mu_{IJ}F^I_{mn}F^{Jmn}
-\frac{i}{4}\nu_{IJ}F^I_{mn}\tilde{F}^{Jmn}
\Big).
\label{4daction1}
\end{align}
The indices $m, n, \cdots$ label the 4d space-time coordinate and $I, J,\cdots=(0,a)$.
A new gauge field strength $F^0_{mn}=2\partial_{[m} A_{n]}$ comes from KK $U(1)$ part.
The dual field strength $\tilde{F}^{Imn}$ is defined by $\tilde{F}^{Imn}=\frac{i}{2}\epsilon^{mnop}F^I_{op}$.
The complex scalar field $z^a$ is given by a combination
\begin{equation}
z^a=a^a+iM^ae^s\mathcal{N}^{-1/3},
\label{moduliz}
\end{equation}
and $G_{a\overline{b}}$, $\nu_{IJ}$ and $\mu_{IJ}$ are expressed  as
\begin{equation}
G_{a\overline{b}}=\frac{e^{-2s}\mathcal{N}^{2/3}}{2}G_{ab},
\end{equation}
\begin{equation}
\nu_{IJ}=\begin{pmatrix}
-\frac{1}{3}c_{cde}a^ca^da^e & \frac{1}{2}c_{acd}a^ca^d
\\
\frac{1}{2}c_{bcd}a^ca^d & -c_{abc}a^c
\end{pmatrix},
\end{equation}
\begin{equation}
\mu_{IJ}=\begin{pmatrix}
e^{3s}+2\mathcal{N}^{2/3}e^sG_{cd}a^ca^d & -2\mathcal{N}^{2/3}e^sG_{ac}a^c
\\
-2\mathcal{N}^{2/3}e^sG_{bc}a^c & 2\mathcal{N}^{2/3}e^sG_{ab}
\end{pmatrix}.
\end{equation}

It is known that this theory is described by the so-called {\it special geometry}.
Let us consider complex scalars $X^I$ of 4d vector multiplets and the prepotential
\begin{equation}
F(X)=-\frac{1}{6}c_{abc}\frac{X^aX^bX^c}{X^0}.
\label{prepotential}
\end{equation}
When we define $F_I(X)=\partial F(X)/\partial X^I$, $F_{IJ}(X)=\partial^2 F(X)/\partial X^I\partial X^J$ and $N_{IJ}=2\textrm{Im}F_{IJ}$, \footnote{We take $X^0$ to be real and $i\left(F_I(X)\overline{X}^I-X^I\overline{F}_I(\overline{X})\right)=1$ as a gauge choice.} and choose 
\begin{equation}
z^{a}=\frac{X^{a}}{X^{0}},\,\,\,\,\,\,\,\,\,\,\,\,\,\,\,
z^{0}=1,\,\,\,\,\,\,\,\,\,\,\,\,\,\,\,
N_{IJ}X^I\overline{X}^J=-1,
\end{equation}
the 4d $N=2$ supergravity coupled to an arbitrary number of  vector multiplets can be described by the physical scalar $z^a$ and the prepotential.
Defining the K{\" a}hler potential $K$ by
\begin{eqnarray}
e^{-K}=-z^{I}N_{IJ}{\overline z}^{J}=\vert X ^{0}\vert ^{-2}, 
\label{eq:kahler}
\end{eqnarray}
we see $G_{a \overline{b}}=\partial _{a}\partial _{\overline{b}}K$ as the K{\" a}hler metric.
The couplings $\nu _{IJ}$ and $\mu _{IJ}$ are also reconstructed in terms of $z^a$, 
\begin{eqnarray}
\nu _{IJ}-i\mu _{IJ}={\overline F}_{IJ}+i\frac{N_{IK}z^{K}N_{JL}z^{L}}{
z^{M}N_{MN}z^{N}},
\end{eqnarray}
through the identification of the scalars (\ref{moduliz}) and the prepotential (\ref{prepotential}). 

At this stage we assume that $z^a$ depends only on the radial coordinate $r$ and take an ansatz 
\begin{eqnarray}
ds^{2}_{(4)}=-e^{2U(r )}dt^{2}+e^{-2U(r )}
\left(dr^{2}+
r^2d\Omega ^{2}_{\text{S}^2}
\right),
\label{eq:metric}
\end{eqnarray}
for the 4d metric.
For electric fields $F^I_{mn}$ and magnetic fields $G_{Imn}\equiv\nu_{IJ}F^J_{mn}-i\mu_{IJ}\tilde{F}^J_{mn}$, we can solve
\begin{eqnarray}
F_{\theta \varphi }^{I}=\frac{p^{I}}{2}{\rm sin}\theta , \:\:\:\:\:
G_{I \theta \varphi }=\frac{q_{I}}{2}{\rm sin}\theta , 
\label{eq:gaugefields}
\end{eqnarray}
where $p^I$ and $q_I$ are magnetic charges and electric charges, respectively, and in type IIA view point $(q_0,q_a,p^a,p^0)$ reads $(\textrm{D0,D2,D4,D6})$-brane charges.
Then a function
\begin{eqnarray}
Z=\frac{e^{K/2}}{2\sqrt{2}}\left (  p^{I}F_{I}(z)-q_{I}z^{I}  \right ),
\label{eq:central}
\end{eqnarray}
gives the central charge of 4d $N=2 $ supergravity theory. 
See appendix \ref{bhpotential} for more detailed calculations.

It is in general known that  solutions satisfying first order equations~\cite{fgk}
\begin{equation}
r^2\frac{d}{dr}U=e^U|Z|,\,\,\,\,\,\,\,\,\,\,
r^2\frac{d}{dr}z^a=e^UG^{a\overline{b}}\overline{\mathcal{D}}_{\overline{b}}\overline{Z}\frac{Z}{|Z|},
\label{BPSeq}
\end{equation}
express  4d BPS black holes, where ${\cal D}_{a}$ is the K{\" a}hler covariant derivative, i.e., 
${\cal D}_{a}Z=\left(\partial _{a}+ \frac{\displaystyle{1}}{
\displaystyle{2}}\partial _{a}K \right)Z$.
At the horizon the scalars $z^a$ are attracted to the values specified by ${\cal D}_{a}Z=0$, so eqs. (\ref{BPSeq}) mean the BPS attractor flow. 
We will argue that they are equivalent to  RG flow equations for the $\textrm{QFT}_2$ holographically dual to the extremal black hole space-time, and eventually derive the non-BPS version of them from this equivalence.


\section{Three Dimensional Gravity as the M5-brane configuration}
\label{3d}

In the rest of this paper we concentrate only on the simplest case  in which decoupled 3d gravity is easily considered, in order to see the correspondence between the extremal black hole  in the bulk  and the corresponding boundary $\text{QFT}_2$.  
Now let us return to the 5d action
\begin{align}
\mathcal{I}_{(5)}=\frac{1}{4\pi^2}&\int 
d^5x
\sqrt{-g_{(5)}}
\Bigg[
R^{(5)}-G_{ab}\partial_M M^a\partial^M M^b
-\frac{\mathcal{N}^{2/3}}{2}G_{ab}
F^a_{MN}F^{bMN}
\Bigg]
+\mathcal{I}_{CS}.
\label{5daction3}
\end{align}
This is slightly different from (\ref{5daction2}).
But it is  useful for the Hamilton-Jacobi formalism in the next section since the coefficient matrix of the kinetic term of $M^a$  has the inverse matrix in contrast to (\ref{5daction2}).
Of course solutions to equations of motion are not affected  if we recall $\mathcal{N}=1$.
As is noted in appendix \ref{correct}, the final results are the same even if we assume to start with the action (\ref{5daction2}).
See appendix \ref{correct} for the more precise prescription.
Technically, to follow these procedures is the most significant point   in finding the beta function and the c-function of the dual theory.


From now on we want to consider the M5-brane configuration, or D0-D4-brane configration in 4d view point.
Since we have already known that this is the static spherically symmetric  black hole in four dimensions, a dimensional reduction on $\text{S}^2$ 
\begin{align}
ds^2_{(5)}&=e^{4\omega}(g^{(3)}_{\mu\nu}dx^\mu dx^\nu)+e^{-2\omega}d\Omega_{\text{S}^2}^2,\notag\\
F^a_{\theta\varphi}&=\frac{p^a}{2}\sin\theta,
\label{5dto3d}
\end{align}
gives the  effective 3d gravity.
The indices $\mu,\nu\cdots$ label the 3d space-time coordinates.
The charge $p^a$ is of the M5-brane wrapping on four-cycles in the $\textrm{CY}_3$.
The  effective 3d Lagrangian is the following:
\begin{equation}
\mathcal{I}_{(3)}=\frac{1}{\pi}\int d^3x\sqrt{-g_{(3)}}
\left[
R^{(3)}-6(\partial \omega)^2-G_{ab}\partial_\mu M^a\partial^\mu M^b-V(\phi)
\right],
\label{3daction}
\end{equation}
where the potential term takes the form\footnote{The  higher derivative generalizations were done in ref.~\cite{hhkt}. }
\begin{equation}
V(\phi)=-2e^{6\omega}+\frac{e^{8\omega}}{4}\mathcal{N}^{2/3}G_{ab}p^ap^b.
\label{potential}
\end{equation}
Here $\phi^A$ denotes the 3d scalars $\phi^A=(\omega,\,M^a)$.
In the next section we will see the connection between this potential and the c-function.

Incidentally, the BPS solution of the M5-brane configuration, or the 5d black ring has been already known.
In the 5d language, the metric and the moduli $M^a$ are found by
\begin{equation}
ds^2_{(5)}
=e^{-s}\left[-e^{2U}dt^2+e^{-2U}(dr^2+r^2d\Omega_{\text{S}^2}^2)\right]+e^{2s}(dy+Jdt)^2,
\end{equation}
together with
\begin{equation}
e^{-4U}=H^3(-H_0),\,\,\,\,\,\,\,\,\,\,
e^{2s}=\frac{-H_0}{H},\,\,\,\,\,\,\,\,\,\,
J=\frac{1}{H_0},
\end{equation}
and
\begin{equation}
M^a=\frac{H^a}{H},
\label{exactm}
\end{equation}
where $H$ means
\begin{equation}
H=\left(\frac{1}{6}c_{abc}H^aH^bH^c\right)^{1/3}.
\end{equation}
They are obtained by harmonic functions
\begin{equation}
H^a=h^a+\frac{p^a}{2r},\,\,\,\,\,\,\,\,\,\,
H_0=h_0+\frac{q_0}{2r}.
\end{equation}
In our notation here, $h_0$ and $q_0$ are both negative.

On the other hand, the 4d solution of the scalars
\begin{equation}
z^a=iH^a\sqrt{\frac{-H_0}{H^3}},
\end{equation}
and the metric (\ref{eq:metric}) with $e^{2U}=H^{3/2}(-H_0)^{1/2}$
are read off from eqs. (\ref{moduliz}) and (\ref{5dto4d}). For simplicity we have assumed that axion fields $a^a$ take zero values.
It can actually be proven that they satisfy the first order equations (\ref{BPSeq}) and that the black hole mass saturates the central charge of $N=2$ supergravity (\ref{eq:central}).

Eq. (\ref{5dto3d}) suggests that the M5 solution can be seen as the asymptotically flat BTZ black hole in three dimensions,
\begin{equation}
ds^2_{(3)}=-r^4\frac{H^3}{-H_0}dt^2+r^4H^6dr^2+r^4H^3(-H_0)\left(dy+\frac{1}{H_0}dt\right)^2,
\label{3dspacetime}
\end{equation}
plus the scalars (\ref{exactm}) and
\begin{equation}
e^{-\omega}=e^{-U-s/2}r=rH.
\label{exactsigma}
\end{equation}
This black hole is extremal, and in the holographic view point it corresponds to the $\textrm{QFT}_2$ in which only left movers are excited.\footnote{At the conformal fixed point corresponding to the horizon, it can be checked that the right moving Virasoro charge $\overline{L}_0$ vanishes.}
In section \ref{attractor} we will mention the extremal non-BPS black hole in which only right movers are excited.

In fact, after taking the near horizon limit, we have
\begin{equation}
ds^2_{(3)}\sim -\frac{p^3}{4(-q_0)}r^2dt^2+\frac{(p^3)^2}{64}\frac{dr^2}{r^2}+\frac{-q_0\,p^3}{16}\left(
dy-\frac{2}{-q_0}rdt
\right)^2,
\end{equation}
where
\begin{equation}
p^3=\frac{1}{6}c_{abc}p^ap^bp^c.
\end{equation}
This space-time has the isometry of $\textrm{AdS}_3$ with a radius $\ell=p^3/4$.
As was shown in~\cite{hhknt}, one can derive the Virasoro algebra for $\textrm{CFT}_2$ on the horizon through the Brown-Henneaux-like computation  with the use of the boundary condition founded by~\cite{ghss}.
The  central charge of the Virasoro algebra is expected to be 
\begin{equation}
c=\frac{3\ell}{2G_\text{N}}.
\label{cir}
\end{equation}
Because in our notation the 3d Newton constant is $G_\text{N}=1/16$, one can obtain a well-known result $c=6p^3$~\cite{msw}.
But in this paper we will not follow this calculation and will concentrate on deriving a more general c-function for the dual $\textrm{QFT}_2$ on each boundary. 

In order to reparametrize the metric (\ref{3dspacetime}) for later convenience,
let us introduce new coordinates
\begin{equation}
\rho=r,\,\,\,\,\,\,\,\,\,\,\tau=\frac{1}{4}(-H_0y+2t)-y
,\,\,\,\,\,\,\,\,\,\,\sigma=\frac{1}{4}(-H_0y+2t)+y.
\end{equation}
Then, the metric (\ref{3dspacetime}) can be transformed into
\begin{align}
ds^2_{(3)}=N^2d\rho^2+\frac{1}{\mu^2}\left[
-(d\tau+N^\tau d\rho)^2+(d\sigma+N^\sigma d\rho)^2
\right],
\end{align}
where
\begin{align}
N=\rho^2H^3,\,\,\,\,\,\,\,\,\,\,N^\tau=N^\sigma=-\frac{q_0}{16\rho^2}(\sigma-\tau),
\end{align}
and
\begin{equation}
\mu^2=\frac{1}{\rho^4H^3}.
\label{scale}
\end{equation}
Notice that the limit $\rho\rightarrow\infty$ and  $\rho\rightarrow0$ leads to $\mu\rightarrow0$ and $\mu\rightarrow\infty$, respectively.
Since $\mu$ gives the length scale of the $\textrm{QFT}_2$,
the spatial infinity of the black hole can be regarded as the UV region, and the horizon does to the IR region.




\section{Holographic RG flow}
\label{rg}

The aim of this section is a direct  check of the gauge/gravity correspondence.
The essence is a relation between  the radial coordinate in gravity  and the scale of  the dual $\text{QFT}$ on the boundary, such as eq. (\ref{scale}).
Then the RG flow which connects  the $\text{QFT}$ on each boundary is understood as the variation of boundary values of the scalars in gravity theory along the radial coordinate.
This is the so-called holographic RG flow, and can be well analyzed by using Hamilton-Jacobi formalism~\cite{dbvv}.
In this section after reviewing this formalism briefly, we find the RG flow equation for the M5-system.

\subsection{Review of Hamilton-Jacobi equation}
\label{reviewofhj}

First, we  decompose the 3d metric into the ``Euclidean" ADM form with respect to the radial coordinate $\rho$, like
\begin{equation}
ds^2=N^2d\rho^2+g_{ij}\left(dx^i+N^id\rho\right)\left(dx^j+N^jd\rho\right). 
\label{ADM}
\end{equation}
Here $x^i$ labels the 2d space-time $(\tau,\sigma)$.
Now we rewrite the kinetic terms of the scalars as
\begin{equation}
-6(\partial \omega)^2-G_{ab}\partial_\mu M^a  \partial^\mu M^b
=-\frac{1}{2}L_{AB}\partial\phi^A \partial\phi^B
\end{equation}
in the action (\ref{3daction}), for simplicity.

Next we insert the ADM decomposition of the metric into the Lagrangian density $L$, and define $\pi^{ij}$ and $\pi_A$ as conjugate momenta of $g_{ij}$ and $\phi^A$ in an usual mannar~\cite{RT}.  
Then, up to total derivative terms, the Hamiltonian density is expressed as
$H= \pi^{ij} \dot g_{ij} + \pi_A \dot \phi^A - L= N \mathcal{H} + N^i \mathcal{P}_i$, 
in which $\mathcal{H}$ and $\mathcal{P}^i$ are defined by
\begin{align}
\frac{1}{\sqrt{-g}} \mathcal{H}&= \frac{1}{(-g)}\left(
(\pi^i_i)^2-(\pi_{ij})^2-\frac{1}{2}L^{AB}\pi_A\pi_B \right)
+V(\phi)-R^{(2)}+\frac{1}{2}L_{AB}\partial_i\phi^A \partial^i\phi^B,
\label{hamiltonian}\\
\frac{1}{\sqrt{-g}} \mathcal{P}^i&= -2\nabla_j\left(\frac{1}{\sqrt{-g}}\pi^{ij}\right)+\frac{1}{\sqrt{-g}}\pi_A\partial^i\phi^A.
\label{momentum}
\end{align}
The inverse matrix $L^{AB}$ is now
\begin{equation}
L^{\omega\omega}=\frac{1}{12},\,\,\,\,\,\,\,\,\,\,\,\,\,\,\,
L^{ab}=\frac{1}{2}G^{ab}=\frac{1}{2}M^aM^b-\mathcal{N} \mathcal{N}^{ab},
\label{inverse}
\end{equation}
where $\mathcal{N}^{ab}$ is defined through $\mathcal{N}_{ab}\mathcal{N}^{bc}=\delta_a^c$.\footnote{In the case of  the action (\ref{5daction2}) we cannot define the inverse matrix $L^{AB}$ like (\ref{inverse}).}
Note that equations of motion of $N$ and $N^i$ imply $\mathcal{H}=\mathcal{P}^i=0$.

Now let $g_{ij}(x)$ and $\phi^A(x)$ be boundary values of the classical solutions at a cut-off scale $\rho_c$.
Substituting the classical solutions into the Lagrangian (\ref{3daction}) and integrating it over the three dimensional space, 
we obtain a functional with respect to $g_{ij}(x)$ and $\phi^A(x)$.
We denote this functional as $S[g,\phi;\rho_c] =
16 \pi G_\text{N} \mathcal{I}_{(3)}$.
As a matter of fact, one can confirm that the functional $S[g,\phi;\rho_c]$ is independent of $\rho_c$, and that boundary values of the conjugate momenta are given by
\begin{align}
  \pi^{ij}(x) = \frac{\delta S}{\delta g_{ij}(x)}, \qquad
  \pi_A(x) = \frac{\delta S}{\delta \phi^A(x)}. \label{eq:mom}
\end{align}
Thus, the Hamilton-Jacobi equation reduces to only two constraints,
\begin{equation}
  \mathcal{H}\left(g_{ij}(x),\phi^A(x),\pi^{ij}(x),\pi_A(x)\right) = 0, \qquad
  \mathcal{P}^i\left(g_{ij}(x),\phi^A(x),\pi^{ij}(x),\pi_A(x)\right) =0,
\end{equation}
with eq.~(\ref{eq:mom}).
The latter constraint implies the invariance under the diffeomorphism of the $\text{QFT}_2$  in $(\tau, \sigma)$ space-time with $\rho$ fixed.
The constraint $\mathcal{H} = 0$ leads to the following equation,
\begin{equation}
  \frac{1}{(\sqrt{-g})^2}\left[-\left(g_{ij}\frac{\delta S}{\delta g_{ij}}\right)^2
  +\left(\frac{\delta S}{\delta g_{ij}}\right)^2
  +\frac{1}{2}L^{AB}\frac{\delta S}{\delta \phi^A}\frac{\delta S}{\delta \phi^B} \right] = V(\phi)-R^{(2)} 
+ \frac{1}{2}L_{AB}\partial_i\phi^A \partial^i\phi^B.
\label{hjeq}
\end{equation}
It is possible from this equation to derive the conformal anomaly for the $\text{CFT}_2$, or the Callan-Symanzik equation for the dual $\text{QFT}_2$ although we will not mention that in this paper.

\subsection{Beta function and c-function of M5-system}
\label{betac}

Before solving the Hamilton-Jacobi equation (\ref{hjeq}), we are required to subtract an UV divergence of the bulk action in the limit $\rho_c\rightarrow \infty$.
For this purpose we divide the functional $S[g,\phi]$ into  local counter-terms and a non-local part $\Gamma[g,\phi]$, 
which is a generating functional with respect to the external sources $g_{ij}(x)$ and $\phi^A(x)$.
Next we assign  weight  $w=0$ to $g_{ij}(x),\,\phi^A(x)$ and $\Gamma[g,\phi]$, 
and $w=1$ to $\partial_i$.
From these assignment and the equation
$\delta\Gamma=\int d^2x(\delta g_{ij}(x)\delta\Gamma/\delta g_{ij}(x)+\delta \phi^A(x)\delta\Gamma/\delta \phi^A(x))$, 
quantities $R^{(2)},\,\delta\Gamma/\delta g_{ij}(x)$ and $\delta\Gamma/\delta \phi^A(x)$ turn out to be $w=2$.

Hence, we assume that the classical action $S[g,\phi]$ is decomposed as 
\begin{equation}
  S[g,\phi] = -\int d^2x\sqrt{-g}\,\big\{ W(\phi) + \cdots \big\} + 16\pi G_\text{N} \Gamma[g,\phi].
\end{equation}
The function of only the scalar fields $W(\phi)$ is the local counter-term with $w=0$.
The dots represent integrands of local counter-terms with $2 < w$.
But $w=2$ terms can be absorbed into $\Gamma[g,\phi]$, and $4<w$ terms are not necessary for the present purpose.
Anyway, substituting this decomposition into eq. (\ref{hjeq}) and comparing the terms with $w=0$, we immediately get
\begin{equation}
  V(\phi)=-\frac{1}{2}W(\phi)^2+\frac{1}{2}L^{AB}\partial_A W(\phi) \partial_B W(\phi),
\label{weight0}
\end{equation}
where $\partial_A$ denotes $\partial/\partial\phi^A$.
Since the potential  in the left hand side is given by eq.~(\ref{potential}) for the present case, 
one can finally obtain
\begin{equation}
W(\phi)=4e^{3\omega}-\frac{1}{2}e^{4\omega}\mathcal{N}^{-2/3}
\mathcal{N}_ap^a,
\label{superpot}
\end{equation}
as a solution to eq. (\ref{weight0}), using the inverse matrix (\ref{inverse}).
Note that at the horizon $(\rho=0)$ this function gives the value  $\left. W(\phi)\right|_{\rho=0}=8/p^3$ from the exact solution (\ref{exactm}) and (\ref{exactsigma}).

From the terms with $w=2$ in eq.~(\ref{hjeq}), we obtain the following relation,
\begin{equation}
  \langle\, T^i_i(x) \,\rangle = -\frac{1}{8\pi G_\text{N}W(\phi)}R^{(2)}
  + \beta^A(\phi)\frac{1}{\sqrt{-g}}\frac{\delta \Gamma}{\delta \phi^A(x)}
  + \frac{1}{16\pi G_\text{N}W(\phi)}L_{AB}\partial_i \phi^A \partial^i \phi^B.
\label{trace}
\end{equation}
Here the energy momentum tensor is defined as
\begin{equation}
  \langle\, T^{ij}(x) \,\rangle=\frac{2}{\sqrt{-g}}\frac{\delta \Gamma[g,\phi]}{\delta g_{ij}(x)}, 
\end{equation}
which is natural from the fact that the constraint $\mathcal{P}^i=0$ means $\nabla_i<T^{ij}(x)>=0$.
And the beta function $\beta^A(\phi)$ is given by
\begin{equation}
  \beta^A(\phi)=\frac{2L^{AB} \partial_B W(\phi)}{W(\phi)}.
 \label{beta}
\end{equation}
Actually, $\beta^A(\phi)$ can be interpreted as the beta function for the dual $\textrm{QFT}_2$ since one can explicitly check that
\begin{align}
\mu\frac{d\omega}{d\mu}&=-\frac{2}{3}
\left(
-1+\frac{e^{3\omega}}{W(\phi)}
\right)
=\frac{2L^{\omega\omega}\partial_\omega W(\phi)}{ W(\phi)}=\beta^\omega(\phi),
\label{betasigma}\\
\mu\frac{dM^a}{d\mu}&=\frac{e^{4\omega}}{W(\phi)}
\left(
p^a-\frac{1}{3}\mathcal{N}_bp^bM^a
\right)
=\frac{2L^{ab}\partial_b W(\phi)}{ W(\phi)}=\beta^a(\phi),
\label{betam}
\end{align}
due to (\ref{exactm}), (\ref{exactsigma}), (\ref{inverse}) and $\mathcal{N}=1$.
They are identified with the beta functions of the dual M5-system.

It is apparent that the third term in the right hand side of eq.~(\ref{trace}) becomes zero since the scalar fields $\phi^A(x)$ are homogeneous on the 2d surface from the explicit solution.
Furthermore, the second term vanishes at the critical point $\rho=0$ because of $\beta(\phi) = 0$.
Thus, we obtain there
\begin{equation}
  \langle\, T^i_i(x) \,\rangle \Big|_{\rho=0}=
  - \frac{1}{24\pi} \frac{3}{G_\text{N} W(\phi)} R^{(2)} \Big|_{\rho=0}. \label{anomaly}
\end{equation}
Vanishing of the beta function indicates that the $\textrm{QFT}_2$ becomes conformally invariant, 
so the above equation corresponds to the conformal anomaly 
for the $\textrm{CFT}_2$ at IR fixed point ($\rho=0$).
From $W|_{\rho=0}=8/p^3$ and $G_{\text{N}}=1/16$, this agrees with the expression of the central charge of the Virasoro algebra (\ref{cir})  in the last section.

At any point $(\rho\neq0)$ the second term of the right hand side in eq. (\ref{trace}) does not vanish, nevertheless a function
\begin{equation}
  \mathcal{C}(\phi) = \frac{3}{G_\text{N}W(\phi)}
 \label{cfunction}
\end{equation}
looks like  the so-called c-function for the dual field theory.
Because the function $W(\phi)$ is always non-negative, 
it is clear that
\begin{equation}
  \mu\frac{d\mathcal{C}(\phi)}{d\mu} = \beta^A(\phi) \partial_A\mathcal{C}=-\frac{3}{2G_\text{N}W(\phi)}\beta^A(\phi)L_{AB}\beta^B(\phi)
  \leq 0.
\end{equation}
The equality is satisfied only at $\rho=0$, where the dual theory becomes conformally invariant and the function (\ref{cfunction}) gives the central charge of the Virasoro algebra (\ref{cir}).
This monotonicity is also  consistent with the Zamolodchikov's c-theorem~\cite{z} although it was in ref.~\cite{gjmt}  proposed that the area of $\text{S}^2$ in 4d black hole view point can be seen as the c-function.
The analysis of the holographic RG flow enables us to say that the function (\ref{cfunction}) with (\ref{superpot}) is also natural as  the c-function of the dual M5-system.

\section{Relation to attractor flow equation}
\label{attractor}


Finally, we discuss relations between the above holographic RG flow  and the attractor flow equations for the extremal black holes.
Setting axion fields to be zero and denoting $z^a=iy^a$, $(y^3)=\frac{1}{6}c_{abc}y^ay^by^c$ and $(yyp)=\frac{1}{6}c_{abc}y^ay^bp^c$, we find
\begin{equation}
Z=\frac{(y^3)^{-1/2}}{8}
\Big(
3(yyp)-q_0
\Big),
\end{equation}
and
\begin{equation}
G^{a\overline{b}}\overline{\mathcal{D}}_{\overline{b}}\overline{Z}
=\frac{i}{2}(y^3)^{1/2}\left(
-p^a+\frac{3(yyp)}{2(y^3)}y^a
+\frac{q_0}{2(y^3)}y^a
\right),
\end{equation}
in the expression of the attractor flow equations (\ref{BPSeq}).

Let us focus on the first one in (\ref{BPSeq}).
It is rewritten as
\begin{equation}
-\frac{1}{4}r^2\frac{d}{dr}\Big(\log (-H_0H^3)\Big)
=\frac{1}{16}c_{abc}H^aH^bp^cH^{-3}-\frac{q_0}{8}(-H_0)^{-1},
\end{equation}
by using the explicit solution of $U$ and $z^a$.
Here we remove only terms involved in $q_0$ in both hand sides of the above equation.
Then it is completely expressed by the 3d fields, like
\begin{equation}
\frac{3}{4\mu^2}W(\phi)-\frac{3}{2\mu^2}W(\phi)\mu\frac{d\omega}{d\mu}=\frac{e^{4\omega}}{8\mu^2}\mathcal{N}_ap^a.
\label{RGatt1}
\end{equation}
This is of course compatible with the RG flow equation (\ref{betasigma}) and (\ref{superpot}).

Next, the second equation in (\ref{BPSeq}) is turned out to be
\begin{align}
&ir^2\frac{d}{dr}\Big(
H^aH^{-3/2}(-H_0)^{-1/2}
\Big)\notag\\
=&\frac{i}{2}H^{-3/2}(-H_0)^{1/2}
\left[
-p^a+\frac{H^{-3}}{4}c_{bcd}H^bH^cp^dH^a+\frac{q_0}{2}(-H_0)^{-1}H^a
\right].
\end{align}
If we remove terms involved in $q_0$ again and devide it by $(-H_0)^{1/2}$, it is also re-interpreted as an equation for the 3d scalars, 
\begin{equation}
\frac{e^{2\omega}}{2\mu^3}W(\phi)
\left(
-\mu\frac{dM^a}{d\mu}+M^a-2M^a\mu\frac{d\omega}{d\mu}
\right)
=\frac{e^{6\omega}}{2\mu^3}\left(
-p^a+\frac{1}{2}\mathcal{N}_bp^bM^a
\right),
\label{RGatt2}
\end{equation}
which is surely consistent  with the RG flow equations (\ref{betasigma}) and (\ref{betam}).

It is, in conclusion, found that the BPS attractor flow equations for the 4d black hole are completly equivalent to
 the RG flow equations for the $\textrm{QFT}_2$ dual to the effective 3d gravity.
For the BPS black holes, the scalars follow the first order flow equation (\ref{BPSeq}) in the bulk, and are fixed at the event horizon.
Even if the values at the spatial infinity, for example $h^a$ and $h_0$ for the D0-D4 case, are chosen arbitrary,\footnote{It is supposed to satisfy the asymptotic flatness.} the values at the horizon are completly determined by the black hole charges.
This fact, that is, the attractor mechanism can be holographically understood by saying that the dual $\textrm{QFT}$ is always attracted at the IR conformal fixed point along the RG flow equation.

As an application, we can derive first order non-BPS attractor   flow equations although only the BPS black hole solution was dealt with so far.
Contrary to the fact that BPS solutions satisfy first order equations, non-BPS black holes, in general, seem to  come from  second order equations of motion. 
It is, however, natural to consider that there are first order flows for non-BPS black holes as well if we expect the existence of the dual $\textrm{QFT}$ satisfying the RG flows on each boundary.

We know that the asymptotically flat BTZ black hole
\begin{equation}
ds^2_{(3)}=-r^4\frac{H^3}{H_0}dt^2+r^4H^6dr^2+r^4H^3H_0\left(dy+\frac{1}{H_0}dt\right)^2,
\label{3dspacetimenonbps}
\end{equation}
\begin{equation}
M^a=\frac{H^a}{H},\,\,\,\,\,\,\,\,\,\,e^{-\omega}=rH,
\label{scalarnonbps}
\end{equation}
where both $h_0$ and $q_0$ are now  positive,
is also a solution, but this is non-BPS because the black hole mass does not saturate the central charge of $N=2$ supergravity (\ref{eq:central}).
It is, nevertheless, straightforward to apply the calculation in the last section to this solution.
One can confirm that (\ref{superpot}) and the flow equations (\ref{betasigma}) and (\ref{betam}) are still the same.
We have checked that the BPS attractor flow equations are consistent with them above, but we can conversely   follow that  calculation.
Starting with eqs. (\ref{RGatt1}) and (\ref{RGatt2}), and using the exact solution (\ref{3dspacetimenonbps}) and (\ref{scalarnonbps}), one arrives at
\begin{equation}
r^2\frac{d}{dr}U=e^U|\mathcal{Z}|,\,\,\,\,\,\,\,\,\,\,
r^2\frac{d}{dr}z^a=e^UG^{a\overline{b}}\overline{\mathcal{D}}_{\overline{b}}\overline{\mathcal{Z}}\frac{\mathcal{Z}}{|\mathcal{Z}|},
\label{nonBPSeq}
\end{equation}
where a new function
\begin{equation}
\mathcal{Z}=\frac{e^{K/2}}{2\sqrt{2}}\left(p^aF_a(z)+q_0\right)
\end{equation}
is defined instead of (\ref{eq:central}).
The moduli are attracted at the point $\mathcal{D}_a\mathcal{Z}=0$, independently of the asymptotic values $h^a,h_0$ .
They are the non-BPS attractor flow equations at least for this $\overline{\text{D0}}$-$\text{D4}$ configuration.

Our approach in the effective 3d gravity is very useful to see the first order flow equation for the non-BPS black hole.
The difference between the BPS (\ref{3dspacetime}) and non-BPS (\ref{3dspacetimenonbps}) solutions is only the sign of the angular momentum of the BTZ black hole.
It is easy  and natural to construct first order flow equations if one starts with the action (\ref{5daction3}) and works in  the canonical formalism (or more precisely, taking into  account of the calculation in section \ref{correct}).
It is an open problem for the future works to write down  such canonical equations with respect to $r$ of more general non-BPS configurations like those discussed in refs.~\cite{cersole,aaot,psrv} or, for instance, the extremal Kerr solution, and to study whether they can be identified with the RG flows.

\section{Conclusion}
\label{conclusion}

In this paper we have extended the claim of the ``Kerr/CFT  correspondence" to the more general gauge/gravity correspondence in the full black hole space-time of the bulk.
For the purpose of  obtaining the decoupled 3d gravity with scalars easily, we presented an example of the M5-system which gives rise to a typical 4d RN black hole. 
Constructing the Hamilton-Jacobi equation in the 3d gravity, we derived the beta function (\ref{betasigma}) and (\ref{betam}) and Zamolodchikov's c-function (\ref{cfunction}).
This indicates that the $\text{QFT}_2$ satisfying the RG flow equations surely lives on each boundary of the extremal black hole.
Moreover, it was confirmed that these holographic RG flow equations are equivalent to the (BPS and non-BPS) attractor flow equations known in four dimensions.
In conclusion, it is found that why the near horizon geometry is $\text{AdS}$ and the attractor mechanism is effective for extremal black holes can be understood from the holographic view point.

Indeed, it is difficult to see such a correspondence for  more general configurations explicitly   since we cannot in general obtain the decoupled 3d gravity. 
And  it is natural to expect the existence of the dual $\text{QFT}_2$  in our example since it is of course a solution of M-theory.
But we claim that for more general  extremal black holes  there also exists  such a  gauge/gravity correspondence in the whole space-time and that the attractor flows are related to the holographic RG flows.

However, we have no idea about what the dual $\text{QFT}_2$ really is although we have proven its existence in this paper.
For the M5-system, it is known that this  reduces to $N=(0,4)$ $\textrm{SCFT}_2$ at the critical point~\cite{msw}, but that is all of our knowledge.
As was shown in the previous section, we have found the exact beta functions (\ref{betasigma}) and  (\ref{betam}) which took the same forms for both left movers and right movers in this two derivative case.
Therefore, more generic behaviors of the dual theory may be investigated by these results obtained from the bulk gravity.
The interpretation of multi-centered solutions, in particular,  seems to be a very intriguing problem as was argued in refs.~\cite{denef, multi}.

The authors of~\cite{witten2} discussed an interesting relation  between the BTZ black hole in 3d gravity and the monster theory as  the $\textrm{CFT}_2$.
On the other hand, there are many $\textrm{QFT}_2$, such as the minimal $\textrm{CFT}$ model, whose deformation and RG flows have been known rather well.
By using the technique used in this paper, we have to investigate what is the true $\textrm{QFT}$ holographically dual to the extremal black hole,  as a next step of the correspondence between the extremal black hole in the bulk and two dimensional field theory on the boundary.


\section*{Acknowledgements}

I wish to thank  my collaborators Yoshifumi Hyakutake, Takahiro Kubota, Takahiro Nishinaka and Hiroaki Tanida for many useful discussions and helpful instructions.
I would also like to thank  Tatsuo Azeyanagi, Wei Li, Keiju Murata, Tatsuma Nishioka, Noriaki Ogawa and Seiji Terashima for useful discussion.
This work is supported in part by JSPS Research Fellowship for Young Scientists.


\appendix

\section{5d superconformal gravity}
\label{superconformal}

It is known that the following action discribes the 5d $N=2$ superconformal gravity with two derivative terms:
\begin{align}
\mathcal{I}_{(5)}=\frac{1}{4\pi^2}\int &
d^5x
\sqrt{-g_{(5)}}
\Bigg[
\frac{1}{4}(\mathcal{N}+3)R^{(5)}+\mathcal{N}_{ab}\left(\frac{1}{2}\partial_M M^a\partial^M M^b+\frac{1}{4}F^a_{MN}F^{bMN}\right)
\notag\\
&
+v^{MN}v_{MN}(3\mathcal{N}+1)+2\mathcal{N}_aF^{a}_{MN}v^{MN}
+\frac{1}{2}D(\mathcal{N}-1)\Bigg]
+\mathcal{I}_{CS}.
\label{5daction1}
\end{align}
The field $D$ is the Lagrange multiplier by which we explicitly impose the constraint $\mathcal{N}=1$ on equations of motion.
The tensor $v_{MN}$ is an auxiliary field as well. 
After redefining the moduli, $M^a\rightarrow \mathcal{N}^{-1/3}M^a$,\footnote{This means $\mathcal{N}\rightarrow1$, which is consistent with the constraint of the original $M^a$.} 
one obtains
\begin{align}
\mathcal{I}_{(5)}=\frac{1}{4\pi^2}\int &
d^5x
\sqrt{-g_{(5)}}
\Bigg[
R^{(5)}-\frac{\mathcal{N}_a\mathcal{N}_b}{3\mathcal{N}^2}\partial_M M^a\partial^M M^b
+\frac{\mathcal{N}_{ab}}{2\mathcal{N}}\partial_M M^a\partial^M M^b\notag\\
&+4v^{MN}v_{MN}+\frac{2\mathcal{N}_a}{\mathcal{N}^{2/3}}F^{a}_{MN}v^{MN}
+\frac{\mathcal{N}_{ab}}{4\mathcal{N}^{1/3}}F^a_{MN}F^{bMN}
\Bigg]
+\mathcal{I}_{CS}.
\label{5daction1}
\end{align}
Moreover, solving the auxiliary field $v_{MN}$ and substituting it into the above, one gets to the action (\ref{5daction2}). 
We can consider that the solution of this new $M^a$ also satisfies $\mathcal{N}=1$.



\section{Black hole potential}
\label{bhpotential}

Assuming 4d RN black holes, we begin with the static metric (\ref{eq:metric}).
In order to solve  equations of motion for $U$ and $z^{a}$ coupled to the 
gauge fields, we put
\begin{eqnarray}
F_{tr}^{I}=\frac{\hat q ^{I}}{2}, \:\:\:\:\:
F_{\theta \varphi }^{I}=\frac{p^{I}}{2}{\rm sin}\theta , \:\:\:\:\:
G_{I t r}=\frac{\hat p _{I}}{2}, \:\:\:\:\:
G_{I \theta \varphi }=\frac{q_{I}}{2}{\rm sin}\theta , 
\label{eq:gaugefields}
\end{eqnarray}
where the magnetic fields are defined as $G_{Imn}=\nu_{IJ}F^J_{mn}-i\mu_{IJ}\tilde{F}^J_{mn}$.
But $\hat q^{I}$ and $\hat p_{I}$ 
are actually given by the  electric and magnetic charges $q_{I}$ and $p^{I}$ of  
the black hole as 
\begin{eqnarray}
& & \hat q^{I}=\frac{e^{2U}}{r^2}[-(\mu ^{-1})^{IJ}\nu _{JK}p^{K}
+(\mu ^{-1})^{IJ}
q_{J}], 
\notag\\
& & \hat p_{I}=\frac{e^{2U}}{r^2}\left[-\nu _{IJ}(\mu ^{-1})^{JK}\nu _{KL}p^{L}+\nu 
_{IJ}(\mu ^{-1})^{JK}q_{K}
-\mu _{IJ}p^{J}\right],
\end{eqnarray}
due to  equations of motion of the gauge fields.
With the gauge field configurations appearing in
(\ref{eq:gaugefields}), the equations of motion 
for the action (\ref{4daction1})
turn out to be 
\begin{eqnarray}
& & U''=e^{2U}V_{BH}, 
\notag
\\
& & -\left \{ U''-2(U')^{2}
\right \}+2G_{a{\bar b}}(z^{a})'
({\overline z}^{b})'-e^{2U}V_{BH}=0, 
\notag
\\
& & \{ G_{a\bar b}({\overline z}^{b})' \}'-\partial _{a}G_{b\bar c}(z^{b})'
({\bar z}^{c})' = e^{2U}\partial _{a}V_{BH},
\label{eq:eqom}
\end{eqnarray}
where $'\equiv d/d(-1/r)$.
The function
\begin{eqnarray}
V_{BH}(z, \bar z, p, q)=\frac{1}{16}(p^{I}, q_{J})
\begin{pmatrix}
(\nu \mu ^{-1}\nu +\mu)_{IK} & -{(\nu \mu ^{-1})_{I}}^{L}
\\
-{(\mu ^{-1}\nu )^{J}}_{K} & (\mu ^{-1})^{JL}
\end{pmatrix}
\begin{pmatrix}
p^{K}
\\
q_{L}
\end{pmatrix}
\label{eq:bhpotential}
\end{eqnarray}
is often called the black hole potential.

Here even if we consider a system with the Lagrangian 
\begin{eqnarray}
{\cal L}(U, z, \bar z )=
(U')^{2}+G_{a{\bar b}}(z^{a})'(\overline{z}^{b})'+e^{2U}
V_{BH}(z, \overline{z}, p, q), 
\label{eq:effectivelagrangian}
\end{eqnarray}
plus a constraint 
\begin{eqnarray}
(U')^{2}
+G_{a{\bar b}}(z^{a})'(\overline{z}^{b})'-e^{2U}V_{BH}
(z, \overline{z}, p, q)=0,
\label{eq:constraint}
\end{eqnarray}
we can also derive the equations of motion (\ref{eq:eqom}).
Thus, this reduced theory is equivalent to the original one (\ref{4daction1}) when the metric and scalars depend only on $r$.

Letting $\mathcal{D}_a$ denote the K\"{a}hler covariant derivative, we here note useful relations
\begin{equation} 
F_I(z)=(\nu_{IJ}-i\mu_{IJ})z^J,\,\,\,\,\,\,\,\,\,\,
\overline{\mathcal{D}}_{\overline{a}}(e^{K/2}\overline{F}_I(z))
=(\nu_{IJ}-i\mu_{IJ})\overline{\mathcal{D}}_{\overline{a}}(e^{K/2}\overline{z}^J),
\end{equation}
and
\begin{equation}
G^{a\overline{a}}\mathcal{D}_a(e^{K/2}z^I)\overline{\mathcal{D}}_{\overline{a}}(e^{K/2}\overline{z}^J)
+e^{K}\overline{z}^Iz^J=\frac{1}{2}(\mu^{-1})^{IJ}.
\end{equation}
By using these, the black hole potential (\ref{eq:bhpotential}) can further be rewritten as~\cite{fk}
\begin{eqnarray}
V_{BH}(z, \bar z, p, q)=\vert Z\vert ^{2} +\vert {\cal D}_{a}Z\vert ^{2}, 
\end{eqnarray}
where 
$Z$ is defined by (\ref{eq:central}) and $\vert {\cal D}_{a}Z\vert ^{2}=G^{a\overline{b}}{\cal D}_{a}Z\overline{\mathcal{D}}_{\overline{b}}\overline{Z}$.
One can verify that the BPS condition (\ref{BPSeq}) is a solution to eq. (\ref{eq:eqom}).
For the particular non-BPS case (\ref{3dspacetimenonbps}) and (\ref{scalarnonbps}),   first order flow equations (\ref{nonBPSeq}) are also a solution.

\section{Precise prescription of Hamilton-Jacobi formalism}
\label{correct}

Begining with the correct $N=2$ action (\ref{5daction2}), one cannot naively construct the Hamiltonian as was done in subsection \ref{reviewofhj}, because the coefficient matrix of the kinetic term of $M^a $ does not have the inverse matrix.
In order to realize the same equations of motion as those from the action (\ref{5daction2}), we recall $\mathcal{N}_a\partial_\mu M^a=\partial_\mu \mathcal{N}$, and introduce a new scalar field $\tilde{\mathcal{N}}$ and a Lagrange multiplier $\lambda$.
Then, after the dimensional reduction on $\text{S}^2$, consider the 3d Hamiltonian density 
\begin{equation} 
H=N\mathcal{H}+N^i\mathcal{P}_i+\lambda (\mathcal{N}-\tilde{\mathcal{N}}),
\end{equation}
where
\begin{align}
\frac{1}{\sqrt{-g}} \mathcal{H}= \frac{1}{(-g)}
\Bigg(
(\pi^i_i)^2-(\pi_{ij})^2&-\frac{1}{2}L^{AB}\pi_A\pi_B \Bigg)
+V(\phi)-R^{(2)}+\frac{1}{2}L_{AB}\partial_i\phi^A \partial^i\phi^B\notag\\
&
-\frac{1}{2(-g)}(-3\tilde{\mathcal{N}}^2)
\pi_{\tilde{\mathcal{N}}}^2
+\frac{1}{2}\left(
\frac{-1}{3\tilde{\mathcal{N}}^2}
\right)\partial_i \tilde{\mathcal{N}} \partial^i \tilde{\mathcal{N}},
\end{align}
\begin{align}
\frac{1}{\sqrt{-g}} \mathcal{P}^i&= -2\nabla_j\left(\frac{1}{\sqrt{-g}}\pi^{ij}\right)+\frac{1}{\sqrt{-g}}\pi_A\partial^i\phi^A+\frac{1}{\sqrt{-g}}\pi_{\tilde{\mathcal{N}}}\partial^i\tilde{\mathcal{N}}.
\end{align}
The matrices $L_{AB}$ and $L^{AB}$ and the potential $V(\phi)$ take the form as before
\begin{align}
L_{\omega\omega}=&12,\,\,\,\,\,\,\,\,\,\,
L_{ab}=2G_{ab}=\left(\frac{\mathcal{N}_a\mathcal{N}_b}{\mathcal{N}^2}-\frac{\mathcal{N}_{ab}}{\mathcal{N}}\right),\notag\\
L^{\omega\omega}=&\frac{1}{12},\,\,\,\,\,\,\,\,\,\,
L^{ab}=\frac{1}{2}G^{ab}=\frac{1}{2}M^aM^b-\mathcal{N}\mathcal{N}^{ab},\notag\\
V(\phi)&=-2e^{6\omega}+\frac{e^{8\omega}}{4}\mathcal{N}^{2/3}G_{ab}p^ap^b.
\end{align}
This Hamiltonian is motivated by a relation 
\begin{equation}
\frac{1}{2}\left(
\frac{2\mathcal{N}_a\mathcal{N}_b }{3\mathcal{N}^2}-\frac{\mathcal{N}_{ab}}{\mathcal{N}}
\right)\partial_\mu M^a\partial^\mu M^b
=
G_{ab}\partial_\mu M^a\partial^\mu M^b
+\frac{1}{2}\left(\frac{-1}{3\mathcal{N}^2}\right)\partial_\mu \mathcal{N}\partial^\mu\mathcal{N}.
\end{equation}
In fact, combining equations of motion for $M^a$, $\pi_a$, $\tilde{N}$, $\pi_{\tilde{\mathcal{N}}}$ and $\lambda$, one can obtain the same 
equation of motion as that of $M^a$ from the original action (\ref{5daction2}).

As we did in subsection \ref{betac}, the equation
\begin{equation}
  V(\phi)=-\frac{1}{2}W^2+\frac{1}{2}L^{AB}\partial_A W \partial_B W+\frac{1}{2}(-3\tilde{N}^2)(\partial_{\tilde{\mathcal{N}}}W)^2,
\end{equation}
is found as a solution of the Hamilton-Jacobi equation with $w=0$.
Therefore, we can easily solve it by a form independent of $\tilde{\mathcal{N}}$,
\begin{equation}
W(\phi)=4e^{3\omega}-\frac{1}{2}e^{4\omega}\mathcal{N}^{-2/3}
\mathcal{N}_ap^a.
\end{equation}
It is checked that the beta function of $\tilde{\mathcal{N}}=1$ of course vanishes and that other beta functions (\ref{betasigma}) and (\ref{betam}) are the same. 
Since the final results are not affected by the terms of $\tilde{\mathcal{N}}$, we temporarily used  (\ref{5daction3}) as if it was the original action.


\end{document}